\documentstyle[amsmath,amssymb,twocolumn,prb,aps,floats,psfig,epsfig]{revtex}

\begin{document}
\twocolumn[\hsize\textwidth\columnwidth\hsize\csname 
@twocolumnfalse\endcsname

\title{Nonadiabatic Superconductivity and Vertex Corrections 
in Uncorrelated Systems} 

\author{P. Paci$^1$,
E. Cappelluti$^2$, C. Grimaldi$^3$ and L. Pietronero$^2$} 

\address{$^1$ Dipartimento di Fisica ``A. Volta'', Universit\'{a} di Pavia, 
Via Bassi 6, 27100 Pavia, Italy \\
and Istituto Nazionale Fisica della Materia, Unit\'a di Roma 1, Italy}
\address{$^2$ Dipartimento di Fisica, Universit\'{a} di Roma 
``La Sapienza", 
Piazzale A.  Moro, 2, 00185 Roma, Italy \\
and Istituto Nazionale Fisica della Materia, Unit\'a di Roma 1, Italy}
\address{$^3$ \'Ecole Polytechnique F\'ed\'erale de Lausanne,
D\'epartement de Microtechnique IPM, CH-1015 Lausanne, Switzerland}

\maketitle 

\begin{abstract}
We investigate the issue of the nonadiabatic superconductivity
in uncorrelated systems.
A local approximation is employed coherently with the weak
dependence on the involved momenta.
Our results show that nonadiabatic vertex corrections are never
negligible,
but lead to a strong suppression of $T_c$ with respect to the conventional
theory. This feature is understood in terms of the momentum-frequency
dependence of the vertex function. In contrast to strongly correlated
systems,
where the small ${\bf q}$-selection probes the
positive part of vertex function, vertex corrections in uncorrelated
systems
are essentially negative resulting in an effective reduction of the
superconducting pairing.
Our analysis shows that vertex corrections in nonadiabatic regime can be
never
disregarded independently of the degree of electronic correlation
in the system.
\end{abstract}

\vskip 2pc]

\narrowtext

One of the characteristics of the high transition
temperature superconductors (HTSC) 
is the small density of charge carriers.\cite{uemura}
As a consequence, the scale of the electronic dynamics, the Fermi energy
$E_F$, is comparable with the typical phonon frequencies 
$\omega_{\rm ph}$ and the 
adiabatic parameter $\omega_{\rm ph}/E_F$ becomes significant.
This situation opens the way to
a scenario in which nonadiabatic effects are relevant.
From a diagrammatic point of view, nonadiabatic effects 
can be taken into account by the inclusion of
vertex corrections arising from the breakdown of 
Migdal's theorem.\cite{migdal} However, in general,
this amounts to consider an infinite set of 
nonadiabatic diagrams whose resummation is a formidable
task. Nevertheless, it is possible to formulate
a perturbative theory by assuming that the order of magnitude
of the first vertex corrections, $\lambda \omega_{\rm ph}/E_F$,\cite{migdal}
is small enough to be treated as an expansion parameter.\cite{psg,gps}
In principle, this assumption is fulfilled by nonadiabatic 
weak coupling systems ($\lambda<1$ and $\omega_{\rm ph}/E_F\sim 1$)
or moderately nonadiabatic strong coupling materials 
($\lambda\sim 1$ and $\omega_{\rm ph}/E_F < 1$).

In previous studies we have shown how such
a perturbative approach accounts for some of the anomalous
properties of the HTSC-materials.
In particular, large values of $T_c$, 
compared with the ones predicted by the Migdal-Eliashberg (ME)
theory, are related in a natural way to the opening
of nonadiabatic channels in the Cooper pairing with no need
of assuming large values of $\lambda$.\cite{gpsprl,cgps}
To understand from a microscopical point of view 
in which way the nonadiabatic channels affect
the superconducting properties, and in particular
the critical temperature $T_c$, a detailed study of  
the momentum-frequency structure of the vertex function is needed.
In fact, vertex function
presents a complex behaviour with respect to the momentum ${\bf q}$ and the 
frequencies $\omega$ of the exchanged phonon.\cite{psg,gps} In particular
the vertex function has been shown to be positive for small values of ${\bf q}$
and negative for large values of ${\bf q}$ compared to $\omega$, 
leading respectively to
an enhancement and to a decrease of the superconducting pairing.
The evaluation {\em a priori} of
the nonadiabatic effects on $T_c$ is thus not at all easy,
since it will depend
on the total balance of negative and positive parts of the vertex function.
In this perspective, specific properties of real materials become
very important, since they can modify the balance favouring or disfavouring
positive or negative parts and determining an enhancement or a suppression of
the critical temperature.

In particular, in strongly correlated systems
the electron correlation due to
onsite Coulomb repulsion has shown to be actually responsible for
a predominance of small-${\bf q}$ scattering 
yielding an effective modulation of the electron-phonon 
coupling.\cite{zeyher,kulic}
In this situation the positive part of
the nonadiabatic vertex corrections is mainly probed, leading to
an net increase of the coupling in Cooper channel
and to a corresponding enhancement of the critical temperature 
$T_c$.\cite{gps,gpsprl,cgps}
On the other hand in uncorrelated systems the 
${\bf q}$-dependence is weak and the negative part of the vertex corrections
at large momenta is expected to lead to a resulting decrease of the 
Cooper pairing and of $T_c$. 

The aim of this short communication is twofold. First, we show that
nonadiabatic effects cannot be neglected also in uncorrelated materials 
(structureless electron-phonon interaction) and vertex corrections play 
a primary role in suppressing the superconducting pairing as long as
$\omega_{\rm ph}/E_F$ is not negligible. Second, we argue that 
the nonadiabatic superconductivity developed by us for the small ${\bf q}$
scattering regime is also a rather good approximation of the
uncorrelated case, 
providing therefore a unified and reasonable description of both the correlated
and uncorrelated cases.

In conventional metals, according to Migdal's theorem,\cite{migdal}
the smallness of the adiabatic pa\-ram\-e\-ter $\omega_{\rm ph}/E_F$
permits to describe successfully the electron-phonon coupled
system by neglecting the vertex corrections in the electron-phonon
interaction.
The application of Migdal's theorem to the superconducting state has
led to the ME equations of superconductivity,\cite{elia} which accurately
describe the properties of conventional superconductors.\cite{carbotte}
A different situation is encountered in nonadiabatic
materials, as we have briefly discussed above, where the breakdown
of Migdal's theorem is expected.
The relevance of the nonadiabatic corrections can be established
by evaluating the vertex function schematized as
\begin{eqnarray}
P(\omega_n,\omega_m;{\bf q})& =& g^2 \sum_{{\bf p},\omega_l} 
D(\omega_n-\omega_l)G({\bf p},\omega_l)
\nonumber\\ 
&&\hspace{9mm}\times G({\bf p+q},\omega_l+\omega_m)\, ,
\label{vertex}
\end{eqnarray}
where $D$
and $G$ represent respectively the phonon and the electron propagators,
$g$ is the electron-phonon matrix element, and
$\omega_n$, $\omega_m$ and $\omega_l$ are fermionic Matsubara frequencies.
We adopt a simple Einstein spectrum with frequency $\omega_0$
for the phonon propagator
$D(\omega_n-\omega_l)=-\omega_0^2/[\omega_0^2+(\omega_n-\omega_l)^2]$.
 
As shown by Migdal in his pioneering work,\cite{migdal}
the vertex function, given in the 
Eq.~(\ref{vertex}), 
scales as $\lambda \omega_{\rm ph}/E_F$ where 
$\lambda= 2 N_0g^2 / \omega_0$ is the dimensionless
electron-phonon coupling and 
$N_0$ is the electronic density of states (DOS) at the Fermi level. 
To obtain this result typical phonon momenta are assumed to be of
the order of the Debye momentum $q_{\rm D}\sim k_F$.
However, as briefly above discussed, this assumption breaks down
in strongly correlated systems where the predominance of
small-${\bf q}$ scattering is important to establish the enhancement 
of the critical temperature. 

The small-${\bf q}$ selection due to strong Coulomb repulsion
in strongly correlated systems can be simulated 
by a cut-off $q_c$ in the exchanged
momentum space so that  
$\sum_{{\bf q}} \rightarrow \sum_{{\bf q}} \theta(q_c-|{\bf q}|)$ 
which restricts the momentum integrations.
In this situation we replace the vertex function 
$P(\omega_n,\omega_m;{\bf q})$
by its average over momenta $P(\omega_n,\omega_m; q_c)$
that depends only on the frequencies and on the cut-off
$q_c$:
\begin{equation}
P(\omega_n,\omega_m;q_c) =
\frac{
\sum_{{\bf q}}  \theta(q_c-|{\bf q}|) P(\omega_n,\omega_m;{\bf q})
}{\sum_{{\bf q}}  \theta(q_c-|{\bf q}|)}.
\label{vmedia}
\end{equation}
We can generalize the ME equations to include the momentum average of
the first order vertex corrections due to the breakdown of
Migdal's theorem.
The generalized ME equations in the nonadiabatic regime
for the self-energy renormalization function $Z$ and 
for the superconducting gap $\Delta$ are the following:\cite{gps}
\begin{eqnarray}
Z(\omega_n)& = &1 + \frac{T_c}{\omega_n}\sum_{\omega_m} 
\Gamma_Z (\omega_n,\omega_m,Q_c)
\eta_m ,
\label{z}\\
Z(\omega_n)\Delta(\omega_n)& =&  T_c\sum_{\omega_m} 
\Gamma_{\Delta} (\omega_n,\omega_m,Q_c) 
\frac{\Delta(\omega_m)}{\omega_m}\eta_m,
\label{gap}
\end{eqnarray}
where $\eta_m = 2\arctan\{E_F/[Z(\omega_m)\,\omega_m]\}$
and $Q_c = q_c/2k_F$ is the dimensionless cut-off.
The kernels of the equations~(\ref{z}) and~(\ref{gap}) are respectively
given by:
\begin{eqnarray}
\Gamma_Z (\omega_n,\omega_m,Q_c)& = &  \lambda D(\omega_n-\omega_m) 
[1+\lambda P(\omega_n,\omega_m,Q_c)],
\nonumber\\
\nonumber\\
\Gamma_\Delta (\omega_n,\omega_m,Q_c) & = & \lambda D(\omega_n-\omega_m) 
[1+2\lambda P(\omega_n,\omega_m,Q_c)]
\nonumber\\
&+& \lambda^2 C(\omega_n,\omega_m,Q_c).
\nonumber
\end{eqnarray}
Explicit expressions of the vertex $P$ and cross $C$ functions 
have been obtained analytically
for small values of $Q_c$ in Refs. \onlinecite{psg,sgp}.
Higher order nonadiabatic corrections in $\Gamma_Z$ and $\Gamma_\Delta$,
here not taken into account,
should be explicitely included in the extreme nonadiabatic regime where
$\lambda P \sim ~1$ ($\lambda C \sim 1$).\cite{dolgov}
This intriguing issue is however beyond the purpose of our paper,
and all the following results apply only when 
$\lambda\omega_{\rm ph}/E_F$ is small enough to permit truncation
of higher order vertex corrections.

The nonadiabatic equations of superconductivity
Eqs.~(\ref{z}) and~(\ref{gap})
have been numerically solved  and the resulting $T_c$ has found to be
strongly enhanced with respect to the adiabatic case
due to the inclusion of vertex corrections and to the presence
of electronic correlation that favours small-${\bf q}$
scattering.\cite{gpsprl,cgps}

A different situation is encountered when we look at
uncorrelated systems where the ${\bf q}$-dependence 
of the relevant physical quantities is weak.
In this case no restriction in the momenta integration is expected
and the momentum average of the vertex function becomes an almost exact
approximation.
As easily seen from Eq.~(\ref{vmedia}) 
this corresponds to a {\em local} theory 
in which the vertex function becomes
\begin{eqnarray}
&&P_{loc}(\omega_n,\omega_m) =
\sum_{{\bf q}}  P(\omega_n,\omega_m;{\bf q})
\nonumber\\
\hspace{5mm} &=&  g^2 \sum_{\omega_l} 
D(\omega_n-\omega_l)G_{loc}(\omega_l)G_{loc}(\omega_l+\omega_m),
\label{localvertex}
\end{eqnarray}
where $G_{loc}$ is the local electron propagator:
\[
G_{loc}(\omega_n)=\int d\epsilon
\frac{N(\epsilon)}{i\omega_nZ(\omega_n)-\epsilon}\, .
\]
For a direct comparison we adopt the same simplifications
of Refs. \onlinecite{psg,gps},
namely we considered a half-filled constant DOS band with
$N(\epsilon)=N_0$, $-E_F \leq \epsilon \leq E_F$.
$E_F$ represents then the Fermi energy.

The nonadiabatic equations of superconductivity for uncorrelated systems
are thus formally obtained by substituting the local vertex function
given by the Eq.~(\ref{localvertex}) in the kernels
$\Gamma_Z$ and $\Gamma_{\Delta}$
of the equations~(\ref{z}) and~(\ref{gap}).
The numerical solution of such equations in local regime 
follows the usual scheme. Therefore, without entering in details,
we are going to discuss the results.
\begin{figure}[t]
\centerline{\epsfig{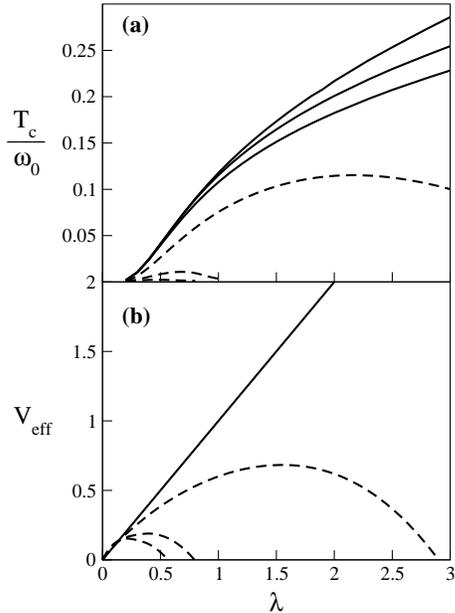}}
\caption{(a) $T_c$ as function of $\lambda$ in uncorrelated
nonadiabatic systems. 
Solid lines correspond to the non crossing
approximation;
dashed lines correspond to the nonadiabatic local theory.
Both the cases are shows (from top to bottom)
 for $\omega_0/E_F= 0.1,\, 0.4,\, 0.7$. 
(b) Behaviour of effective pairing interaction 
$V_{\rm eff}$ 
as function of $\lambda$ in non crossing approximation (solid line) and 
in the nonadiabatic local theory (dashed lines).}
\label{fig1}
\end{figure}

In Fig.~\ref{fig1}a we show
the superconducting transition
temperature $T_c$ of uncorrelated nonadiabatic systems
as function of dimensionless
electron-phonon coupling $\lambda$ for three different
adiabatic parameters.
To evidence the effects of the vertex corrections,
the results of the nonadiabatic local theory are
compared with those obtained by the conventional one
for values of $\lambda$ up to the unphysically large $\lambda=3$,
where of course the perturbative approach breaks down
($\lambda\,\omega_0/E_F \sim 1$).
Solid lines represent the ME solutions
where nonadiabatic effects are just included
by considering finite energy bandwidth. This is thus
equivalent to a non crossing approximation (NCA).
Dashed lines are the numerical results of 
Eqs. (\ref{z})-(\ref{gap}), 
where the vertex
function is given in the local theory by Eq.~(\ref{localvertex}).
From top to the bottom, different lines correspond to
adiabatic parameters: $\omega_0/E_F = 0.1,\,0.4,\,0.7$.

As shown in Fig.~\ref{fig1}a nonadiabatic effects in
uncorrelated systems lead to a
drastic reduction of the critical temperature $T_c$
with respect to the conventional theory.
This result confirms the above qualitative discussion
suggesting that in uncorrelated systems the negative part
of the vertex function is more relevant yielding therefore an effective
reduction of the pairing. The increase of this effect by increasing
both the electron-phonon coupling $\lambda$
and the adiabatic parameter $\omega_0/E_F$ seems also to point
towards a similar conclusion, since the vertex corrections scale
as $\lambda \omega_0/E_F$.

In order to quantify this concept we parametrize the magnitude of
the ``effective'' superconducting pairing, $V_{\rm eff}$,
with the static limit of the superconductive kernel: 
$V_{\rm eff} = \Gamma_{\Delta}(\omega_n=0,\omega_m=0)$.
In conventional ME theory $V_{\rm eff}$ reduces to the simply
bare electron-phonon coupling constant $V_{\rm eff}=\lambda$,
while the opening of nonadiabatic channels strongly affects $V_{\rm eff}$
through the vertex function calculated in the local theory. 
The dependence
of the effective interaction $V_{\rm eff}$
on the bare electron-phonon coupling 
$\lambda$ in both the cases is plotted in 
Fig.~\ref{fig1}b. The similarity between the behaviours of
$T_c$ and of $V_{\rm eff}$ is striking pointing out that
this particular limit of the superconductive kernel
and of the vertex function is directly reflected on the critical
temperature $T_c$: the static negative limit\cite{psg} of the vertex function
reduces $V_{\rm eff}$ and induces a strong supression of $T_c$.
The non monotonic behaviour of $T_c$ in uncorrelated
nonadiabatic systems is thus simply related to the non monotonic underlying
behaviour of $V_{\rm eff}$: the maximum is determined roughly by the value 
of $\lambda$ which gives $d V_{\rm eff} / d \lambda = 0$ corresponding to
 $4 \lambda P_{loc}(\omega_n=0,\omega_m=0)= -1$
(cross function is negligible and can be omitted for the discussion).
We conclude that nonadiabatic effects are dominant also
in uncorrelated systems where they lead to a strong reduction of $T_c$.

Let us address now whether the nonadiabatic theory developed in 
Ref.~\onlinecite{gps,gpsprl} for small momentum transfers\
(small $Q_c$) are capable 
of reproducing the results of the local theory ($Q_c=1$).
In fact, it would be interesting to define a common approach which
would permit to span from uncorrelated to correlated cases.
From this point of view, the small-$Q_c$ expansion
of the vertex function used in Refs.~\onlinecite{psg,gps}
appears a promising tool
to extrapolate to intermediate $Q_c$'s.
In particular, as we can see from Eq.~(\ref{vmedia}),
the uncorrelated case corresponds in this framework to
a $Q_c=1$. We stress that the equivalence between this procedure
and the local theory is only formal since using $Q_c=1$ in a 
small-$Q_c$ expansion is obviously an approximation.
\begin{figure}[t]
\centerline{\epsfig{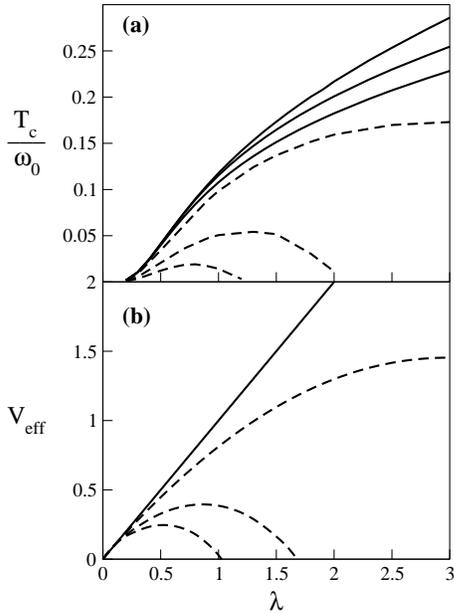}}
\caption{$T_c$ (panel a) and $V_{\rm eff}$ (panel b) 
calculated
by the small-${\bf q}$ expansion of the vertex corrections for $Q_c=1$. 
Solid and dashed lines 
as defined in the previous caption.}
\label{fig2}
\end{figure}

In Fig.~\ref{fig2} the critical temperature $T_c$ and
the ``effective'' superconducting pairing $V_{\rm eff}$
calculated within the small-${\bf q}$ approach with $Q_c=1$ are shown.
The overall features of both $T_c$ and $V_{\rm eff}$ are quite the same
as in the local theory
with a slight overestimation of $T_c$ and $V_{\rm eff}$. This is
not unexpected since the small-${\bf q}$ expansion emphasizes
the positive region of the vertex function and as a consequence
underestimates the net negative magnitude of it.
However the qualitative agreement between Figs.~\ref{fig1} and \ref{fig2}
is quite good suggesting that the nonadiabatic evaluation
of the vertex function based on the small-${\bf q}$ expansion
can actually interpolate from weak to large correlation.
\begin{figure}[t]
\centerline{\epsfig{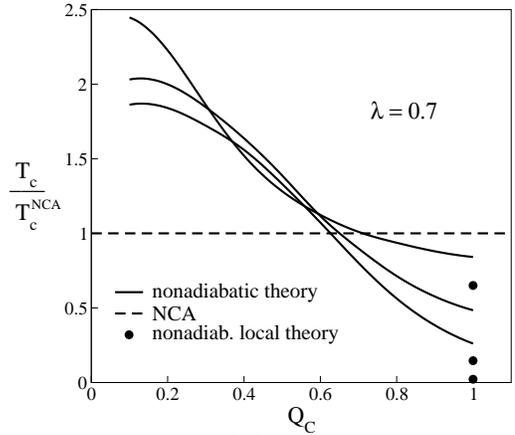}}
\caption{$T_c$ as function of $Q_c$ as evaluated by
the small-${\bf q}$ expansion (solid lines). Dashed lines represents
$T_c$ in non crossing approximation and filled circles the
nonadiabatic local theory for uncorrelated systems.
From top to bottom:
$\omega_0/E_F = 0.1, 0.4, 0.7$}
\label{fig3}
\end{figure}
This is confirmed in Fig.~\ref{fig3} where we plot
the evolution of the critical temperature $T_c$ in nonadiabatic regime
calculated within the small-${\bf q}$ approach (solid lines)
as function of $Q_c$. In an exact theory the solid lines would
end at $Q_c=1$ in the filled circles, representing the nonadiabatic
local theory.
The discrepancy is shown to be quite small and almost independent
of the nonadiabatic parameter. In general, we can identify 
two different regimes: the first one, for small $Q_c$'s, is
relevant for strong correlated
systems where the critical temperature is effectively enhanced with respect
the conventional theory by nonadiabatic effects; the second one which
represent weakly correlated compounds where nonadiabatic vertex corrections
produces a significant reduction of $T_c$.
While the first regime is expected to be well described by
the small-${\bf q}$ approach, Fig.~\ref{fig3} shows
that is works qualitatively well also in uncorrelated systems providing
therefore a unique tool to evaluate the nonadiabatic effects on
the critical temperature. This appears even more important since
the analytical study of the vertex function in small-${\bf q}$ expansion
allows to understand from a microscopic point of view the way
through the which the nonadiabatic vertex function enhances or reduces
$T_c$ respectively in correlated and uncorrelated systems.

In conclusion, we have investigated the nonadiabatic effects
in uncorrelated systems. We have shown that the breakdown of Migdal's
theorem and the consequent inclusion of vertex diagrams lead to a strong
suppression of the critical temperature $T_c$.
Nonadiabatic effects can not be neglected in uncorrelated materials
as well as in correlated ones when the Fermi energy is comparable
to the phonon frequencies. We show also that the nonadiabatic theory
based on a small-${\bf q}$ expansion, early introduced in previous papers,
works quite well even in uncorreleted systems where no predominance
of forward scattering is present. Of course, as already stressed above, 
this conclusion holds true as long as higher order vertex corrections
can be neglected ($\lambda\omega_{\rm ph}/E_F$ sufficiently small).

As a final observation, we would like to remark that the present analysis
has been restricted to the half-filling case. Away from half-filling,
and in particular for Fermi levels very close to the bottom (top)
of the band,
the vertex function changes significantly its structure becoming
roughly shapeless in momentum space and mainly positive.\cite{perali} 
In such a situation
nonadiabatic vertex corrections can give rise to an enhancement of $T_c$
even for uncorrelated systems.\cite{freericks}
This case can be for instance relevant for
the recently discovered superconductivity at $T_c \simeq 40$ K
in MgB$_2$,\cite{akimitsu} where the chemical potential is very close to the top
of the $\sigma$-bands, with $E_F \simeq 0.5$ eV, thus comparable 
to the phonon frequencies.

\end{document}